 \documentstyle[11pt]{article}

\title{\vspace{4cm}\large\bf
New Appearance of Cosmic Coincidence Problem
}
\author{Arthur D.~Chernin\\
 Sternberg Astronomical Institute, Moscow University, Moscow, 119899,
Russia,\\
Astronomy Division, FIN-90014 University of Oulu, Finland,\\
and Tuorla Observatory, University of Turku, Piikki\"o, FIN-21500, Finland\\
}

\date{~}

\begin{document}

\maketitle

\begin{abstract}
\noindent
The integrals of the Friedmann cosmology equations are identified as
constant physical characteristics for both vacuum and non-vacuum
cosmic energies. The integrals are found to be numerically
coincident. A model shows that the coincidence can naturally originate
in the elecroweak scale physics at TeV temperatures. The coincidence
seems to reflect a new type of symmetry that relates vacuum and
non-vacuum energies.

PACS numbers: 98.80.Cq

\end{abstract}
%
%
{\em Introduction. ---} Recent observations suggest that the
cosmological expansion is accelerated [1,2]; this is found in the
brightness-redshift distribution of high-redshift
type-Ia supernovae [1] and confirmed by the bulk of all
observational evidence that come from the cosmic age, large scale
structure, and most convincingly from the combination of the
cosmic microwave background (CMB) anisotropy and cluster dynamics [2,3].
It is assumed that cosmic acceleration is produced by cosmological vacuum,
which has constant energy density
$\rho_V \simeq 2 \times 10^{-123} M_{Pl}^4$,
where $M_{Pl} \simeq 1 \times 10^{19}$ GeV is the Planck energy scale.

In the Friedmann models, both energy density and pressure control
cosmological expansion, so that the sum $\rho + 3 p$ presents the
effective gravitating density in an isotropic universe. When the sum is
negative, gravity becomes anti-gravity and accelerates the expansion.
It is the case for vacuum, since vacuum has positive density
$\rho_V$ and negative pressure $p_V = - \rho_V$.
Vacuum accelerates the expansion, if it dominates over `ordinary' cosmic
energies at the present era. The vacuum density measured
in the units of the critical density is $\Omega_V = 0.7 \pm 0.1$, while the
dark energy density is $\Omega_D = 0.3 \pm 0.1$, the baryon density is
$\Omega_B = 0.02 \pm 0.01$, and the ultra-relativistic
energy (CMB photons, gravitons, etc.) is
$\Omega_R = 0.6 \alpha \times 10^{-4}$ (where $1 < \alpha < 30-100$ is a
dimensionless factor that accounts for possible non-CMB contributions to
the energy) [2,3].

The density of vacuum is indeed larger than
the total density of the non-vacuum energies. But
surprisingly enough, the difference of the densities is not too big,
and all the four are rather close to each other, especially the densities
of vacuum and dark energy. It may mean that the vacuum started to dominate
over the other energies not long ago, and so the present era is the
era of transition from dark energy domination to vacuum domination. But why
are all the four major densities  coincident now? What is the physical
nature of this phenomenon? This is the `cosmic coincidence problem'
 that poses a severe challenge to current cosmological concepts
 (see [2,3] and references therein).

One possible approach to the problem would be to assume that
the acceleration of the expansion is produced by non-vacuum energy  which
has negative pressure and negative effective gravitating
density [4]. This energy form is called quintessence;
it can naturally be realized in some scalar field models in which the field
depends on time, and so the density of quintessence is diluted with
the  expansion. If so,  the densities involved  in the observed cosmic
coincidence are all diluted, which makes them seemingly more similar
to each other. Quintessence density might temporarily or even always be comparable
to the rest of the energy densities [5]. Although such an idea may make the
closeness of all cosmic densities natural, it does not explain the
coincidence that the quintessence field becomes settled with a finite energy
density comparable to the matter energy density just now (see, for instance,
[6]).

Another prospective on the cosmic coincidence problem is suggested by a model
of Ref. [6] which proposes that the constant vacuum density may be determined
in terms of two fundamental energy scales: the `reduced' Planck scale
$ m_{Pl} \sim 10^{18}$ GeV and the electroweak scale $M_{EW} \sim 1$ TeV.
Under some special assumptions, the model [6] leads to
$\rho_V \sim (M_{EW}^2/m_{Pl})^4$. The numerical value of this combination
of the energy scales is rather close,
on the order of magnitude, to the observed figure [1,2].

In the present Letter, a framework is suggested that is
alternative to the idea of quintessence.
The cosmic acceleration is understood below as a dynamical effect of vacuum
with constant density. Moreover, instead of time-dependent densities,
constant time-independent numbers are introduced to the framework and
identified as genuine physical characteristics of non-vacuum energies.
The constants are among the basic quantities in cosmology; they are
integrals of the Friedmann differential equations. For both non-relativistic
and ultra-relativistic energies, the integrals express the conservation of
the total number of particles in the comoving volume. It is interesting 
that the same equations attribute a special integral to vacuum as well, 
considering vacuum on a common basis with ordinary non-vacuum energies.

Two results are reported in the Letter. The first one comes from an empirical 
analysis of the problem. The evaluation of the cosmic integrals is made
for vacuum, dark energy, baryons and ultra-relativistic energy with the
use of the observed cosmic densities. It is found that the integrals
are near-coincident numerically. Thus the cosmic coincidence
appears in a novel version:  this is now a coincidence of constant
time-independent numbers (in contrast to the original version of the
problem and its treatment in terms of of quintessence). If this is not
to be a numerical accident, the coincidence of the integrals looks like
a new type of basic and stable regularities (a symmetry) in the evolving 
universe. Anyway, the integral coincidence exists as long as the
energies themselves exist.

The second result comes from an attempt to clarify the physical nature of 
the integral coincidence on the grounds of fundamental physics. A simple 
kinetics theory model is developed
for the freeze-out  of non-relativistic relic particles in the early
universe; this is a well-know mechanism of the `origin of species'   
in cosmology (see texts of Ref.[11]). The integrals 
enter the model naturally as its basic quantities. To specify the
underlying  physics, it is assumed that 1) the vacuum density is
determined in terms of the electroweak energy scale and the Planck scale,
as is proposed by the field theoretic model [6];
2) the same two fundamental energy scales determine also the epoch of the
freeze-out, so that the redshift at this epoch is $z \sim m_{Pl}/M_{EW}$;
with this $z$, the temperature is $T \sim M_{EW} \sim 1$ TeV at the
freeze-out. These assumptions are consistent with the cental role 
played by electroweak physics in early cosmic evolution [11]. 
Under the two assumptions, the model demonstrates explicitly that the 
integrals for vacuum, non-relativistic and ultra-relativistic energies 
are coincident and have the correct numerical value. Thus the 
discovered coincidence of the cosmic integrals turns out to be  a natural 
outcome of the freeze-out kinetics mediated by the electroweak scale physics 
at the cosmic age  $ \sim 10^{-12}$ sec.

{\em Integrals. ---}
The Friedmann cosmology equations,
\begin{equation}
a^{-2} (a')^2 = \kappa \rho a^2 - E,  \;\;\;
d \rho = - 3 (\rho + p) d \ln a,
\end{equation}
\noindent
relate density and pressure to $a$, which is
the curvature radius (E = +1, -1) or a scale factor ($E = 0$);
the prime denotes a derivative with respect to conformal time,
$\kappa = 8\pi G/3 = (8\pi/3) M_{Pl}^{-2}, E = -1, 0, +1 $,
in open, spatially flat and close models, respectively.
With the Hubble constant $h_{100} = 0.65 \pm 0.15$, the observed densities
lead to either an open model or a spatially flat
model (see [3] and references therein).

For any cosmic energy having  equation of state $p = w \rho$ with a constant
parameter $w$ , the integral of the second (`thermodynamical') equation
in Eq.(1) is $C = \rho a^{3(1+w)} = Const > 0$.
With this relation, the first (`mechanical') equation of Eq.(1) gives
$a$ as a function of time; in the resulting solution,
integral $C$ appears in a combination with the gravitational constant:
\begin{equation}
A = [ (\frac{1 + 3w}{2})^{2} \; \kappa C]^{\frac{1}{1 + 3w}} =
[ (\frac{1 + 3w}{2})^{2} \; \kappa \rho a^{3(1+w)}]^{\frac{1}{1 + 3w}}.
\end{equation}
\noindent
In this form, integrals for different $w$  have
the same dimension (the length) and so can be compared with each other.

For the present epoch of accelerating expansion, the solution is
\begin{equation}
a(t) = A_V f(t), \;\;\; f(t) = \sinh (t/A_V), \;\; \exp (t/A_V),
\;\; \cosh (t/A_V),
\end{equation}
\noindent
for $E = -1, 0, +1$, respectively. Here $A_V = (\kappa \rho _V)^{-1/2}$,
according to Eq.(2) with $w = -1$; the scale factor (for $E = 0$) is
normalized to $A_V$. The acceleration is attributed to vacuum alone.
The Hubble constant is $H \sim A_V^{-1} $, for each $E$ soon
after the transition to vacuum domination; the
current cosmic age is $t_0 \sim H^{-1}$ in the transition era. As a result,
one has approximately $a(t_0) \sim A_V$ for all the three models.

The integrals for dark (D) energy, baryons (B) and ultra-relativistic (R)
energy  with $w = 0, 0, 1/3$,
respectively, enter corresponding cosmological solutions in the
same quite natural way -- see, for instance, an exact
solution for both non-relativistic and ultra-relativistic energies [8].

The value of each of the four integrals can be found with the respective
observed density and  $ a(t) = a(t_0)$; this way one has:
$ A_V = (\kappa \rho _V)^{-1/2} \sim 10^{61} M_{PL}^{-1},
A_D = \frac{1}{4} \kappa \rho_D a^3 \sim10^{60} M_{Pl}^{-1},
A_B = \frac{1}{4} \kappa \rho_B a^3 \sim 10^{59} M_{Pl}^{-1},
A_R = (\kappa \rho _R)^{1/2} a^2  \sim 10^{59} M_{Pl}^{-1}$.
In the estimation of $A_R$, a conservative value
$\alpha = 1$ is adopted which accounts for the CMB only.
(Note that the integral for quintessence with $w = - 2/3$, for example, is
$A_{Q} \sim 10^{61} M_{Pl}^{-1}$, if it is estimated with
 $\rho_Q(t_0) = \rho_V$. A special case
 $w = - 1/3$  is not included in Eq.(2); but using directly the
 corresponding solution of the Friedmann equations, one finds:
 $A_{-1/3} = A_V$.)

Thus the four cosmic integrals prove to be close to each other within
two orders of magnitude:
\begin{equation}
A_w  \sim 10^{60 \pm 1 } M_{Pl}^{-1}, \;\;\; w = [-1, 0, 0, 1/3].
\end{equation}
This is a novel version of the cosmic
coincidence, which appears now as a tetramerous constant coincidence of
constant numbers (in contrast to the temporary coincidence of
the densities, in the original version of the problem).
The coincidence of the two integrals
$A_B$ and $A_R$ was recognized soon after the discovery of the CMB [9].

The integrals exist in the universe since the epoch at which the respective
forms of energy start to exist themselves, and the integrals will exist
until the decay of the protons at $t \sim 10^{31-32}$ sec and/or the decay of
the particles of dark matter. Despite evolution of the ratios of the
densities over many orders of magnitude,
the constant integrals are the same and coincident
during all this exraordirily large time interval.

As for the observed coincidence of the densities, it is clear
that the densities of vacuum and dark matter
are coincident, because this is the era of
transition to the dynamical dominance of vacuum over dark energy,
as is said above. The question of why we
happen to live in such  a special era may  be
treated  on the ground of antropic principle [10].
The coincidence of all the four major densities now is seen as
a direct consequence of  the basic relation of Eq.(4) and the
temporary (and accidental, in this sense) coincidence $a(t_0) \sim A_A$
at the present epoch.

{\em Model. ---} Is the observed  coincidence of cosmic integrals a
simple empirical fact or a reflection of a deeper physical regularity
in nature? In a search for an answer to this question, one should
address the physics of the early universe.
To start with, one may
consider D- and B-energies together as a non-relativistic thermal relic of
early cosmic evolution. Then, one may consider the kinetics of freeze-out
process in the early universe (see standard texts [11] and also a recent
work [6]). As is well-known,
for stable (or long-living) particles of the mass $m$, the abundance freezes
out when the temperature $T$ falls down the mass $m$, and the expansion rate
$1/t$ wins over the annihilation rate, $\sigma n$ (here the annihilation
cross-section $\sigma \sim m^{-2}$). At that particular moment, the particle
density is $n \sim 1/(\sigma t) \sim m^2 (G \rho_R)^{1/2}$.
Using Eq.(2) for $A_R$ and
for $A_M = A_D + A_B$ with $\rho_M \sim m n$, one may now find how the two
non-vacuum integrals are related in the model:
\begin{equation}
A_M \sim a m^3 M_{Pl}^{-2} A_R.
\end{equation}
\noindent
One also has at that moment $\rho_R \sim m^4$, and because of this
\begin{equation}
 A_R \sim  a^2 m^2 M_{Pl}^{-1}.
\end{equation}
Here and in Eq.(5) $a \simeq A_V (1+z)^{-1}$, and $z$ is the redshift
at the freeze-out epoch;
in this way, the vacuum integral comes to the model.

To specify the underlying fundamental physics, one may
refer again to the standard texts [11] that point out a special significance
of electroweak physics in the early universe. If so, it is natural to
identify the mass $m$ of Eqs.(5,6) with the electroweak energy scale
$M_{EW} \sim 1$ TeV. Then, one may exploite the field theoretic model [6]
and adopt its major result mentioned above, i.e. the relation for the
vacuum density in terms of the two fundamental energy sacles, $M_{EW}$
and $M_{Pl}$: $\rho_V \sim (M_{EW}/M_{Pl})^8 m_{Pl}^4$. The only difference
here from [6] is that the standard Planck scale is used, not the reduced
Planck scale (see below). With this density, the vacuum
integral is
\begin{equation}
A_V \sim  (M_{Pl}/m)^4 M_{PL}^{-1}.
\end{equation}

Arguing along this line, one may expect that the epoch of the freeze-out is
determined by the electroweak scale physics as well. If so,
the redshift $z$ at the freeze-out epoch may be
a simple combination of the same two masses:
\begin{equation}
z \sim M_{Pl}/m.
\end{equation}

Now the kinetics model is described by a system of four algebraic Eqs.(5-8)
(with $m = M_{EW}$) for the four numbers $A_M, A_R, A_V, z$. The solution
of the system is:
\begin{equation}
A_M \sim A_R \sim A_V \sim  (M_{Pl}/m)^4 M_{PL}^{-1}.
\end{equation}
\noindent
Thus the coincidence of the cosmic integrals appears as a natural result
of the freeze-out process mediated by electroweak physics.
This physics determines also the value of the integrals.

Following [6] and another recent work [12],  one may introduce to Eq.(9) the
gravitational scale $M_G$, or the reduced Planck
scale $m_{Pl}$, instead of the standard Planck scale:
$M_G \simeq m_{Pl} \simeq g M_{Pl}$,
where $g \simeq 0.1-0.3$. The constant factor $g$ accounts for the fact
that the
gravity constant $G$  enters in  combinations like $ 8 \pi G/3,  6\pi G$,
or  $32 \pi G/3$  exact formulas of cosmology. (Similarily, a
few dimensionless factors, like the effective number of degrees of freedom,
etc., may also be included in the model  -- see again
Ref.[11]). So one gets finally:
\begin{equation}
 A_w \sim g^4 (M_{Pl}/M_{EW})^4 M_{Pl}^{-1} \sim 10^{61 \pm 1} M_{Pl}^{-1},
\;\;\; w = [-1, 0, 1/3].
\end{equation}
\noindent
A quantitative agreement with the empirical result of Eq.(4) seems
satisfactory here.
With $A_V$ of Eq.(10), one finds that the vacuum density is
$\rho_A \sim g^8 (M_{Pl}/M{EW})^8 M_{Pl}^4 \sim 10^{-122 \pm 2} M_{Pl}^4$,
which is also in good agreement with the observed value [1,2]
$\rho_A \sim 10^{-123} M_{Pl}^{-1}$.

Note that the numerical value of the redshift in Eq. (8)
$z \sim g M_G/M_{EW} \sim 10^{15}$, and so the
temperature at the freeze-out is $T \sim 1$ GeV $\sim M_{EW}$, which
is consistent with the central role of the electroweak scale
in the model above.

{\em Symmetry. ---}
The tiny energy density of vacuum  poses the most serious naturalness problem
in theoretical physics [7]. With the identification of cosmic
integrals as genuine physical quantities for both vacuum and non-vacuum
energies, the naturalness problem finds its natural non-vacuum counterparts.
In the framework suggested above, all the forms of cosmic energy can be 
treated in an unified manner and on the common ground of the cosmic integrals. So the
naturalness problem is transformed, it looses its uniqueness and gets its 
place in a
broader context that includes dark energy, radiation, etc. at the same rate. A more general problem is this:  
Why are $A_V \sim A_D \sim A_B \sim A_R \sim 10^{60} M_{Pl}^{-1}$?

Instead of the naturalness problem, one faces a new fundamental problem,  
which is the problem of symmetry that relates vacuum and non-vacuum forms 
of energy.
The symmetry is described by Eqs.(4,10), in terms of
cosmic integrals. This is translational
symmetry in the one-dimensional space of energy forms, or of the
values of the parameter  $w$.
Only three points of the space are directly observed at the phenomenological
level: $w = -1, 0, 1/3$. The two more were mentioned above:
$w =  -2/3, -1/3$; each of them has a clear physical  interpretation.
The Zeldovich super-stiff equation of state,  $w = 1$, may also be included to
this collection. Quintessence is associated with a segment
$-1 < w < - 1/3$. The symmetry is not  perfect, it turns out to be
inexact, at a level of a few percent, on relative logarithmic scale.

Is the space of energy forms discrete or continuous? Is it really finite
$-1 \le w \le 1$? Is the nature of the symmetry due
completely to electroweak physics?  Are quark-hadron physics, GUT, SUSY,
etc.  also mixed into making the symmetry inexact? These are the questions
that are invited by the new appearance of the cosmic coincidence problem,
and the list is obviously far incomplete.

The work was partly supported by the grant of the Academy of Finland.
\vspace{2cm}


[1] S. Perlmutter {\it et al.}, Astrophys. J {\bf 517}, 565 (1998);
A.G. Riess {\it et al.}, Astron. J {\bf 116}, 1009 (1998).

[2] J. Cohn, astro-ph/9807128; S. Carol, astro-ph/0004075; N.Bahcall,
J. Ostriker, S. Perlmutter, and P. Steinhard, Science {\bf 284}, 1481 (1999);
O. Lahav, P.Lilje, J. Primack, and M. Rees, MN RAS {\bf 251}, 126 (1991);
K.K.S. Wu, O. Lahav, and M.J. Rees, Nature (London) {\bf 397}, 225 (1999).

[3] L. Wang, R.R. Caldwell, J.P. Ostriker, and P.J. Steinhardt, Astrophis. J
{\bf 530}, 17 (2000).

[4] R.R. Caldwell and P.J. Steinhardt, Phys.Rev. {\bf D 57}, 6057 (1998); 
 B. Ratra and P.J.E. Peebles, Phys. Rev. D {\bf 37}, 3406 (1988)
 J.P. Ostraiker and P.J. Steinhardt, Nature (London) {\bf 377}, 600 (1995);

[5] I. Zlatev, L. Wang, P.J Steihard, Phys. Rev. Lett. {\bf 82}, 896 (1999); 
A. Hebecker and C. Wetterich, Phys. Rev. Lett. {\bf 85}, 3339 (2000); V.A.
Rubakov, Phys. Rev. D {\bf 61}, 061501 (2000); A.D. Dolgov, Phys. Rev. D
{\bf 55}, 5881 (1997); C. Armandariz-Picon, V. Mukhanov, and P.J. Steinhard,
Phys. Rev. Lett. {\bf 85}, 4438 (2000); I.G. Dymnikova, Phys. Lett.
B {\bf 472}, 33 (2000);
S. Dodelso nand  M. Kaplinghat,
E. Steward, Phys. Rev. Lett. {\bf 85}, 5276 (2000); A.Albrecht and C.
Skordis, Phys. Rev. Lett. {\bf 84}, 2076 (2000).

[6] N. Arkani-Hamed, L.J. Hall, Ch. Kolda, H. Murayama,
Phys. Rev. Lett. {\bf 85}, 4434 (2000).

[7] S. Weinberg, Rev. Mod. Phys. {\bf 61}, 1 (1989).

[8] A.D. Chernin, Astron. Zh. {\bf 42}, 1124 (1965); Engl. transl. Sov.
Astron. -AJ {\bf 9}, 871 (1966).

[9] A.D. Chernin, Pisma ZhETF {\bf 8}, 633 (1968); Engl. transl. Sov.
Phys. - JETP Lett. {\bf 8}, 391 (1968);  A.D. Chernin, Nature (London)
{\bf 220}, 250 (1968) - beware of typos.

[10] S. Weinberg, Phys. Rev. Lett. {\bf 59}, 2607 (1987).

[11] Ya.B. Zeldovich, I.D. Novikov, The Structure and Evolution of the
Universe. The Univ. Chicago Press, Chicago and London (1983); A.D.
Dolgov, Ya.B. Zeldovich, M.V. Sazhin, Cosmology of the Early Universe
(in Russian). Moscow Univ. Press, Moscow (1988); E.W. Kolb, M.S. Turner,
The Early Universe. Addison-Wesley Pubs., New York (1990).

[12] M.Kawasaki, M. Yamaguchi, T. Yanagida, Phys. Rev. Lett. {\bf 85},
3572 (2000).

\end{document}